\documentclass[]{aa}
\usepackage{graphicx}

\usepackage{natbib}
\bibpunct{(}{)}{;}{a}{}{,}

\begin{document}
\title{From Tidal Dwarf Galaxies to Satellite Galaxies}
\author{F. Bournaud \inst{1} and P.-A. Duc \inst{2}}
\offprints{F. Bournaud \email{Frederic.Bournaud@obspm.fr}}
\institute{Observatoire de Paris, LERMA, 61 Av. de l'Observatoire, F-75014, Paris, France
\and CNRS UMR~7158, AIM, \& CEA/DSM/DAPNIA, Service d'Astrophysique, Saclay, 91191 Gif-sur-Yvette Cedex, France
}

\date{Received 22 March 2006 / Accepted 10 May 2006}

\abstract{
The current popular cosmological models have granted to the population of dwarf satellite galaxies a key role: their number, location and masses constrain both the distribution of dark matter and the physical evolution of their hosts. In the past years, there has been increasing observational evidence that objects with masses of dwarf galaxies can form in the tidal tails of colliding galaxies and speculations that they could become satellite-like galaxies orbiting around their progenitors and thus be cosmologically important. Yet, whether the so-called ''Tidal Dwarf Galaxies'' (TDGs) candidates are really long-lived objects and not transient features only present in young interacting systems is still largely an open question to which numerical simulations may give precise answers. We present here a set of 96 N-body simulations of colliding galaxies with various mass ratios and encounter geometries, including gas dynamics and star formation. We study the formation and long-term evolution of their TDG candidates. Among the 593 substructures initially identified in tidal tails, about 75\% fall back onto their progenitor or are disrupted in a few $10^8$ years. The remaining 25\% become long-lived bound objects that typically survive more than 2~Gyr with masses above $10^8$~M$_{\sun}$. These long-lived, satellite--like objects, are found to form in massive gaseous accumulations originally located in the outer most regions of the tidal tails. Studying the statistical properties of the simulated TDGs, we infer several basic properties that dwarf galaxies should meet to have a possible tidal origin and apply these criteria to the Local Group dwarfs. We further found that the presence of TDGs would foster the anisotropy observed in the distribution of classical satellite galaxies around their host. Identifying the conditions fulfilled by interacting systems that were able to form long-lived tidal dwarfs -- a spiral merging with a galaxy between 1/4 and 8 times its mass, on a prograde orbit, with an orbital plane inclined up to 40 degrees to the disk plane -- and estimating their fraction as a function of redshift, we roughly estimate their contribution to the overall population of dwarfs. We conclude that a small but significant fraction of them -- a few percent -- could be of tidal origin. This number may be underestimated in particular environments such as the vicinity of early type galaxies or in groups. 

}
\authorrunning{Bournaud \& Duc}
\titlerunning{}
\maketitle


\section{Introduction}

The properties of dwarf satellite galaxies that surround spirals and ellipticals are often used to constrain cosmological models. Their number has been actively debated since cosmological simulations have shown that the low-mass dark halos are much more numerous around their massive hosts than the dwarf galaxies identified around the Milky Way \citep[e.g.][]{klypin99,moore99}. Moreover, the spatial distribution of satellites can provide information on the mass and shape of dark haloes \citep{zaritsky,brainerd05,LK06}. Finally, they play, as building blocks, a major role on the physical evolution of massive galaxies with which they will eventually merge. It is generally assumed in such studies that dwarf galaxies have a cosmological primordial origin and were not formed recently. However, there is growing evidence that objects with masses typical of dwarf galaxies form in the tidal tails that surround colliding and merging spiral galaxies \citep[e.g.][]{DM94,DM98,hibbard01,mendes,temporin03,knierman}. Many of these so-called Tidal Dwarf Galaxies (TDGs) appear to be self-gravitating objects \citep{braine,bournaud04}. If these tidal objects are long-lived, they could contribute to the total population of dwarf satellites in addition to primordial dwarfs. Their statistical properties would then be modified and the possible constraints on cosmological models would have to be updated. 
 
Dwarf galaxies presently observed to form in tidal tails could however be short-lived. Indeed dynamical friction may cause a rapid orbital decay. \citet{hibbard95} have shown that a large part of the once expelled tidal material falls back onto the parent spiral galaxies in a few hundreds of Myr, but the time-scale increases to Gyr in the outer regions; thus the tidal dwarfs formed there may have a significant lifetime. Also, dwarf galaxies may be disrupted by the tidal field of their progenitor even if the internal orbits of their stars are not resonant with their orbital period \citep[e.g.][]{fleck}; this disruption process can take a few billion years depending on their mass, orbit, and concentration. Thus, the lifetime of tidal dwarfs is a priori rather uncertain and their cosmological importance far from being proven.

Actually, no real old Tidal Dwarf galaxies, still surviving after the merger of their progenitors and the vanishing of the umbilical cord linking them to their parents, has yet been unambiguously found although candidates were discussed in the literature \citep[][and references therein]{hunter,duc04iau}. Because of the difficulties to identify a tidal origin in evolved galaxies, numerical simulations appear to be unavoidable to predict their number. Sofar, numerical modeling has mainly been used to study the formation of objects in tidal tails. The gravitational collapse of pre-existing clouds or regions of tidal tails \citep{elmegreen,BH92} can lead to the formation of clumps along tidal tails with typical masses of $10^{6-8}$~M$_{\sun}$. According to \citet{wetz05} and \citet{wetz06}, a massive gaseous component is required to achieve this collapse. More massive accumulations of matter can form in the outer regions of tidal tails when dark halos are assumed to extend much further than the optical radius of galaxies \citep{bournaud03}. Their formation is initiated by a kinematical mechanism, and self-gravity can make these accumulations of matter collapse into dense star-forming objects \citep[][hereafter D04]{duc04}. Simulations assuming massive and extended gaseous disks in the colliding spirals produce objects that seem to become satellites galaxies \citep[e.g.][D04]{cox} but the conditions for their survival have not yet been studied in detail. The fate of classical infalling satellites has been studied in numerous simulations \citep[e.g.][]{mayer1,mayer2,mayer3}. In such models, the dwarfs were assumed to contain a somehow "shielding" layer of dark matter, while tidal dwarfs do not in a CDM scenario \citep{BH92}. \citet{kroupa97} studied the survival of tidal-like dark-matter free dwarf satellites but started his simulations once the galaxies had already been formed and hence could not address the role of their initial distribution on their evolution. He argued that the objects produced in his model -- with final masses of only $10^5$~M$_{\sun}$ -- become long-lived after having lost up to 99\% of their initial mass. These results cannot be easily extended to more massive objects since the latter undergo more dynamical friction and may have a different morphology. Hence, self-consistent simulations of the formation and subsequent evolution/survival of TDGs are still required to estimate the contribution of such objects to the population of dwarf satellite galaxies.

In this paper, we study the long-term survival of massive dwarf-like objects formed in tidal tails. A set of 96 numerical simulations of galactic encounters were carried out, varying the geometrical parameters of the collision and the relative masses of the colliding galaxies. They are used to predict the number and distribution of long-lived -- with life expectancies greater than one billion year -- tidal objects that will eventually look-like classical dwarf satellites. The numerical techniques and parameters are described in Sect~2. In Sect.~3, we detail our results on the formation criteria and survival time of the tidal objects; we then determine their characteristics once they have become long-lived satellite galaxies. In Sect.~4, we compare them with the properties of respectively the currently forming TDG candidates observed in young interacting systems and the satellites around massive hosts such as those identified in the Sloan Digital Sky Survey and in the Local Group. In particular we discuss rough calculations on the expected fraction of dwarfs of tidal origin among all classical ones. Our conclusions are summarized in Sect.~5.


\section{Numerical simulations}

\subsection{Numerical techniques and parameters}
The simulations presented in this paper use the same particle-mesh FFT code as in D04. In particuler, a sticky-particles scheme is used to model the ISM\footnote{D04 explained that the modeling of the ISM is not critical for the formation of massive tidal objects, and indeed \citet{cox} have obtained the formation of comparable objects with an SPH code} and star formation is described by a local Schmidt Law. Initial conditions for disk galaxies and dark haloes are largely similar to those described in D04. Galaxies are assumed to contain 15\% of gas (gas-to-visible mass ratio, including gas outside the stellar disk). Each disk galaxy is embedded in an extended halo that maintains a flat rotation curve up to ten times the stellar disk radius, consistent with cosmological predictions. 

We use a 512$^3$ cartesian grid to compute the gravitational potential, with a gravitational softening of 380~pc. The most massive galaxy is modeled by 10$^6$ particles for each component (stars, gas, dark matter). Its stellar mass is $2 \times 10^{11}$~M$_{\sun}$ and it contains 10\% of gas. Its stellar disk radius is 15~kpc and its gaseous disk radius 45~kpc -- HI is indeed observed much beyond the optical radius in spiral galaxies \citep[e.g.][]{RH94}. The dark halo is modeled by a pseudo-isothermal sphere, with an asymptotic velocity of 205~km~s$^{-1}$ and a core radius of 10~kpc. The circular velocity at the dark halo virial radius (280~kpc) is 220~km~s$^{-1}$. All the timescales discussed in this paper are hence realistic for galaxies similar to the Milky-Way and must be rescaled accordingly for other galactic mass/virial velocities. The number of particles in the companion depends on its mass; its size is scaled by the square root of its mass to keep the stellar disk surface density constant.

Ninety-six galaxy mergers with various mass ratios and various orbital parameters have been simulated. The detailed parameters for each run are given in Appendix~\ref{appen}. The range of values explored for each parameter is summarized in Table~\ref{params}. In order to increase the statistical size of our sample of TDG candidates, we have privileged cases favorable to the formation of long-lived tidal dwarfs. For instance, because retrograde orbits are found not to form long-lived TDGs, we have mainly simulated prograde cases. Of course, we take into account this bias in our statistical study.

The simulations have been run until $2$~Gyr after the first pericenter of the relative orbit of the two galaxies. The ''merger'' typically occurs 50--300~Myr after this pericenter. We were actually 
able to follow the evolution of the objects formed in tidal tails for more than 1.5~Gyr after their formation which is enough to know whether they have become long-lived dwarf satellites, i.e. objects orbiting around their massive host for a few dynamical times.

\begin{table*}
\centering
\begin{tabular}{lcc}
\hline
\hline
Parameter & simulated range & range most favorable to long-lived TDG formation \\
\hline
Relative velocity & 50 to 320~km~s$^{-1}$ & 50 to 250~km~s$^{-1}$ \\
Impact parameter & 15 to 200 kpc & 30 to 200 kpc \\
Orbit inclination & 0 to 60 degrees & 0 to 40 degrees \\
Orbit orientation & Prograde/Retrograde & Prograde only \\
Mass ratio & 10:1 to 1:10 & 4:1 to 1:8 \\
\hline
\end{tabular}
\caption{Summary of the explored ranges for orbital and mass parameters, and indication of the range favorable to the formation of long-lived TDG. The detailed parameters, and number of long-lived TDGs formed in each simulation, are given in Appendix~\ref{appen}. The relative velocity of the colliding galaxies is computed for an infinite distance, neglecting dynamical friction before the beginning of the simulation. The orbit inclination for each galaxy is the angle between the disk plane and the orbital plane. The mass ratio is given as the TDG-progenitor galaxy to tidally disturbing galaxy mass ratio.}\label{params}
\end{table*}

\subsection{Detection of massive substructures in tidal tails}\label{crit}

We analyzed the spatial distribution of matter at different epochs of the simulations. We identified the massive substructures along the tidal tails that may become long-lived "Tidal Dwarf Galaxies". The tidal objects were sought in the gas component plus the young stars formed in situ after the periaster. Old stars from the parent's disks were not considered as they seem to play a minor role in the formation of structures in tidal tails, at least in the most recent scenarii put forward: the kinematical pilling-up of gas at the tip of a tail (D04) or the dynamical formation of gravitational clumps in massive gaseous tidal tails \citep{wetz06}.

The criterion for an ''object'' to be detected is that the mass of gas and young stars within a diameter $d$ must be larger than a threshold $M_0$. Two combinations of mass threshold / maximal diameter have been used : 
$$M_0 = 3\times 10^8 \mathrm{\;within\;}d<6 \mathrm{\;kpc}$$
or 
$$M_0 = 1\times 10^8 \mathrm{\;within\;}d<3 \mathrm{\;kpc}$$
We do not use the combination $$M_0 = 1\times 10^8 \mathrm{\;within\;}d<6 \mathrm{\;kpc}$$ since it was sometimes found to include parts of tidal tails that visually do not correspond to any substructure. Objects of $10^8$~M$_{\sun}$ contain $2 \times 10^3$ to $10^4$ particles, depending of their gas fraction, so that objects above our mass threshold contain enough particles to be well resolved. 

The identification of objects was made around pixels with a density larger than $\mu_0=M_g/(\pi d^2/4)$, and the center of the spherical boundary (the 6 or 3 kpc diameter) was computed to maximize the included mass. At this stage, one object may in fact correspond to two different condensations of matter. We decided that one object must be separated in two if two components more massive than the mass threshold and separated by a density $\mu < \mu_0$ were found. Note that we have rarely observed objects that were first considered as double to merge into a single one, so that the somehow arbitrary threshold used to separate them was realistic.

The mass center of an object, used to determine its ''position'', was defined as the mass center of pixels with a density larger than $\mu_0$ within the object external radius. We only considered objects located outside 1.5 times the initial stellar disk radius of each parent spiral galaxy, in order not to include spiral arms or other features belonging to the outer disk. Objects located at smaller radii would anyway not survive for a few $10^8$~yr before merging and do not deserve being considered as potential TDGs. In all the paper, $t=0$ corresponds to the first pericenter between the two parent spiral galaxies, not to the beginning of the simulations. The criterion for the detection of massive substructures in tidal tails was first applied at $t=200$~Myr. Later on, every 100~Myr up to $t=2$~Gyr, each simulation was re-analyzed to know whether detected objects were still present (i.e., they were still obeying our mass/size and radial distance criterion) and identify new ones that were not formed at earlier times. 
 
\begin{figure*}[!h]
\centering
\includegraphics[width=14cm]{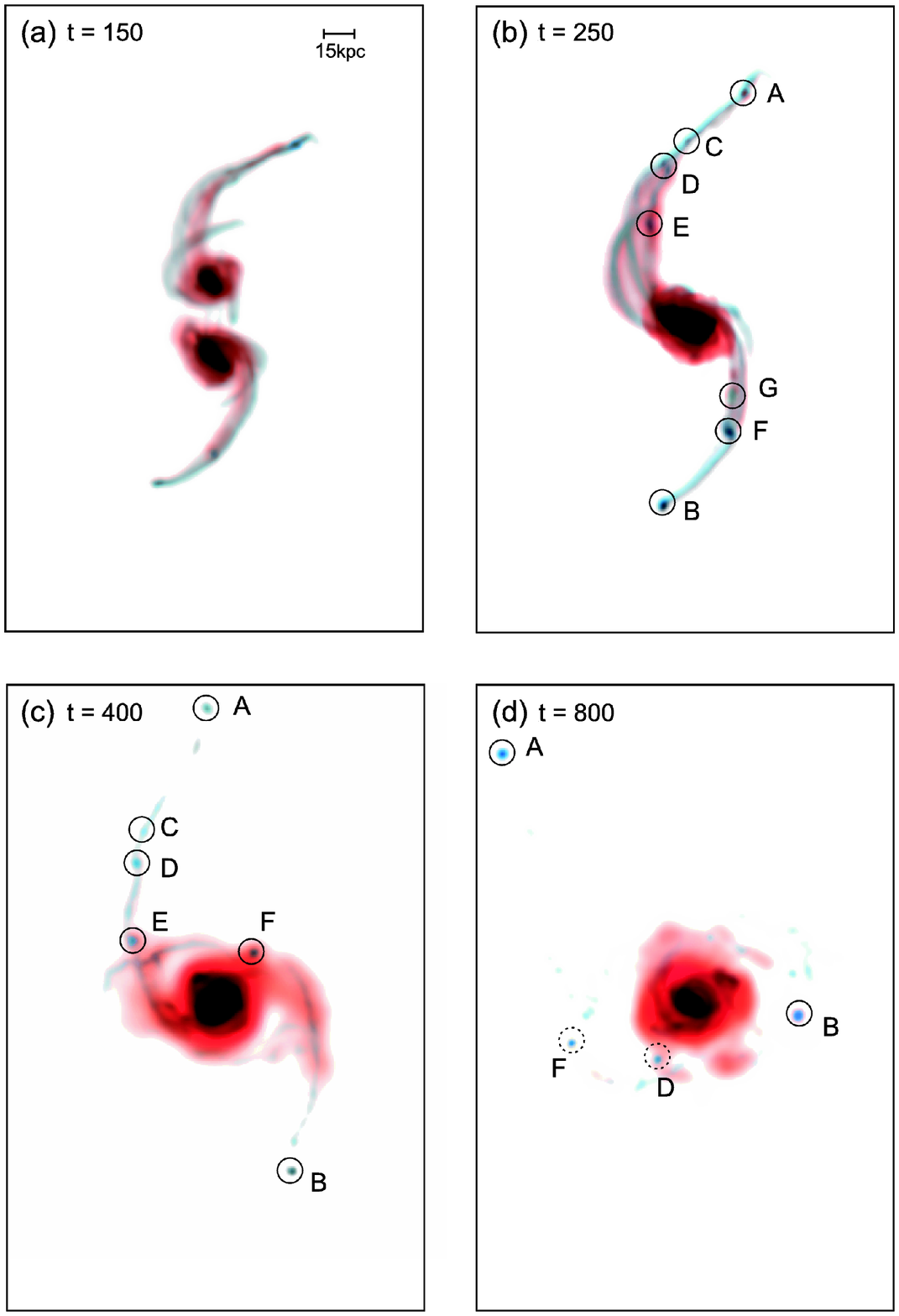}
\caption{Face-on snapshots for run~69 (coplanar encounter). Red and blue colors correspond to regions dominated by old stars vs. gas and young stars. The system is shown at four different instants : (a) at $t=150$~Myr after the pericenter, gas accumulation progenitors of TDGs A and B are already visible at the tip of the tails -- (b) Location of the 7 TDG-candidates more massive than $10^8$~M$_{\sun}$. Initial masses (in $10^8$~M$_{\sun}$) A:~14, B:~8, C:~1.5, D:~2.4, E:~7, F:~6, G:~1.2. -- (c) at $t=400$, object G has fallen back onto the progenitors, and objects D, C, E, F begin to fall-back, while objects A and B remain at larger radii. -- (d) at $t=800$, only objects A and B still exist according to our criterion. Remnants of structures F and D are visible, but they have lost much of their mass: they now have masses of 0.6 and 0.8 respectively (object F has passed within 20~kpc from the merger remnant center). TDGs A and B, located at larger radii, did not undergo such a mass loss: they now have masses of 12 and 6 respectively. Object B falls back onto the merger remnant at $t=1.7$~Gyr; object A survives up to the end of the simulation at $t=2$~Gyr.}\label{ex_substructure}
\end{figure*}

An example of the detection of sub-structures in a simulated merger, using our criteria, is shown in Fig.~\ref{ex_substructure}.

Our detection algorithm based on a density threshold might be considered rather basic  compared to
 more sophisticated algorithms, like the friend-to-friend one. Another approach could also have been 
 to select only the gravitationally bound structures that were likely to become long-lived dynamically 
 independent objects. However, the goal of our simulations is mainly to make relevant comparison with observations. Projection effects and a limited resolution make the identification of bound sub-structures in real systems difficult (see \citet{bournaud04} and \citet{HB04}). Most of the so-called TDG candidates described in the literature were actually identified based on their shape rather than on kinematical data. Because we wished to treat our simulations as would be real images, we decided to use a criterion solely based on the apparent morphology. We thus first included some unbound objects, like observations do, but following their evolution, we could a posteriori reject them. Beside, this allowed us to study the process of tidal disruption and the falling back of tidal material.


\section{Results: statistics on the simulated merging systems}

\subsection{Number, lifetime and masses of tidal sub-structures}

Over our sample of 96 simulations (184 parent spiral galaxies), we identified 593 massive substructures in tidal tails, i.e. an average of 3.2 tidal substructure per parent galaxy. 423 of these structures were first detected at $t=200$, and most the other ones at $t=300$ or $t=400$. 
 
At $t=500$~Myr, 207 objects -- about one third of all the detected structures -- still exist. Conversely, the two other thirds have disappeared in less than 300~Myr. At $t=1$~Gyr, 143 objects are still present, and 119 at $t=2$~Gyr (20\% of the total). 63\% of the vanishing objects have fallen back onto their parent galaxies; the remaining objects ''disappeared'' because their mass dropped below our $10^8$~M$_{\sun}$ threshold. This can be due to tidal disruption or to the fact that they were not gravitationally bound.

\begin{figure}[!h]
\centering
\includegraphics[width=8cm]{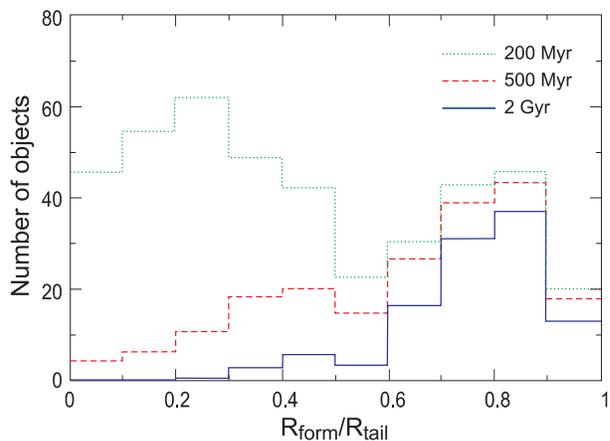}
\caption{Number of objects as a function of their {\it formation} radius, defined as the radius at which they are first detected, normalized by the total radial extent of the tidal tail -- the tidal tail extent is defined to contain 98\% of the mass that has been pulled out 1.5 times the initial gaseous disk radius. The results are shown for the objects existing at $t=200$ (green dotted), 500 (red dashed), and 2000~Myr (blue solid). The longest-lived objects are those formed near the tip of tidal tails, many of which still exist after 2~Gyr. On the other hand, gravitational clumps and other tidal debris formed at smaller radii rarely survive more than 500~Myr. }\label{dist_rad}
\end{figure}

The distribution of their radius of formation (i.e. the radius at which they are first detected) along the tidal tails is shown on Fig.~\ref{dist_rad}. It is roughly bimodal, with two thirds of the objects formed within the first half of the radial extent of the tail, while the other third is formed near the tip of the tails. A close analysis of the simulations (e.g., Fig.~\ref{ex_substructure}) shows that the first kind of objects correspond to loose unbound structures and to gravitational clumps formed all along the tails (such as the ones described by \citet{BH92} or \citet{wetz06}), while the second kind of objects correspond to massive accumulations of matter formed near the tip of tails according to the mechanism described by D04.

\begin{figure}[!h]
\centering
\includegraphics[width=8cm]{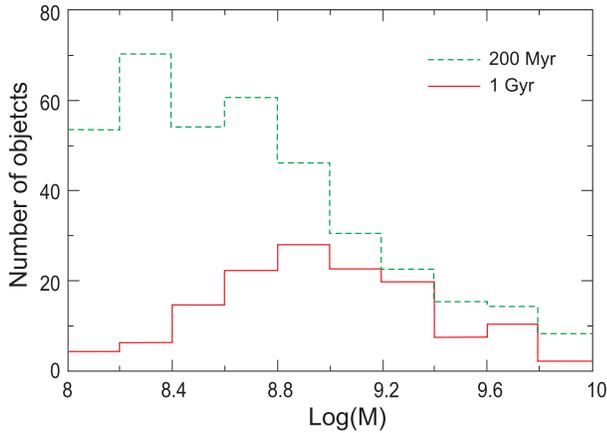}
\caption{Mass spectrum of the objects detected in tidal tails at $t=200$~Myr (green dashed) and $t=1$~Gyr (red solid). The most massive objects that have the longest life are formed near the tip of tidal tails.}\label{histo_mass}.
\end{figure}

Each type of tidal objects appear to have different lifetimes:
\begin{itemize}
\item self-gravitating accumulations of material formed near the tip of tidal tails generally survive for several $10^8$~yr or a few Gyr
\item other objects formed at smaller radii tend to disappear in less than 1~Gyr, and even often in only 500~Myr.
\end{itemize}
Both categories of objects also show differences in their respective masses, as shown in Fig.~\ref{histo_mass}. Most structures detected at $t=200$ have a mass smaller than $10^9$~M$_{\sun}$; but those surviving at least $t=1$~Gyr have the highest masses, with 45\% of them exceeding $10^9$ solar masses. 
 
Therefore, according to our simulations, only the objects formed at the tip of tidal tails can become long-lived, massive, dwarf satellite galaxies. The less massive objects that condensed at smaller radii along the tails tend to fall back rapidly on their parent galaxies or rapidly lose most of their mass, probably contributing to the population of super stellar clusters or even globular clusters observed around merger remnants. Our simulations do not have the resolution to study them in details.

\subsection{Constraints for the formation of long-lived tidal objects}

We now focus on the ''long-lived'' objects able to survive at least 1~Gyr -- those preferentially formed in the outermost regions of interacting systems. The table in Appendix~\ref{appen} indicates whether such objects are formed for each individual run and each parent galaxy. Table~\ref{params} summarizes the constraints obtained for the individual parameters of the collision (eventhough they are not fully independent). The three main restrictions can be summarized as follow: a given spiral galaxy can form long-lived tidal objects provided that:
\begin{itemize}
\item the orbit is {\it prograde}: the angle $\theta$ between the disturbed galactic disk spin and the companion orbital spin 
(see Figure~8 in \citet{duc2000}), is smaller than 90 degrees. 
\item its orbital plane is inclined by {\it less than 40 degrees}
\item the mass of the companion is at least one fourth of that of the target galaxy (otherwise the tidal field cannot drive matter far enough from the parent galaxy) and less than 8 times the target galaxy mass (otherwise tidal objects may form but rapidly fall onto the very massive companion). In other words, for {\it mass ratios in the} 1:1--8:1 {\it range}, at least one of the two spiral galaxies can form long-lived objects. For mass ratios in the 1:1--4:1 range, long-lived objects can form in the material of both 
galaxies.
\end{itemize}
When these three conditions are achieved, most mergers will lead to the formation of long-lived tidal substructures, except those with small impact parameters (typically below 30~kpc), but the latter condition is a minor constraint since the small impact parameters are statistically far less frequent.

When a spiral galaxy merges with a companion inside the parameter range described above, according to the results given in Table~\ref{appen}, on the average 1.9 objects are still visible at $t=500$, and 1.3 survive at least 1~Gyr.

\subsection{Statistical properties of dwarf galaxy satellites of tidal origin}

\subsubsection{Spatial 3D distribution}

\begin{figure}[!h]
\centering
\includegraphics[width=8cm]{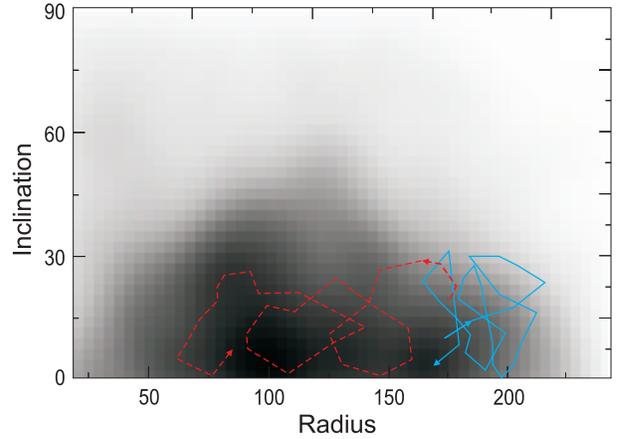}
\caption{Distribution of the long-lived TDGs (surviving at least 1~Gyr) in the radius--inclination plane. The radius is measured from the progenitor galaxy center; the inclination is the latitude in the progenitor disk frame. Data are accumulated from 500 to 2000~Myr. The orbital tracks of two objects are shown. They have the same initial radius, but one (shown in red dashed) was formed on a more eccentric orbit attested by a larger radial excursion; it undergoes a larger radial decay.}\label{points_r_i}
\end{figure}

\begin{figure}[!h]
\centering
\includegraphics[width=8cm]{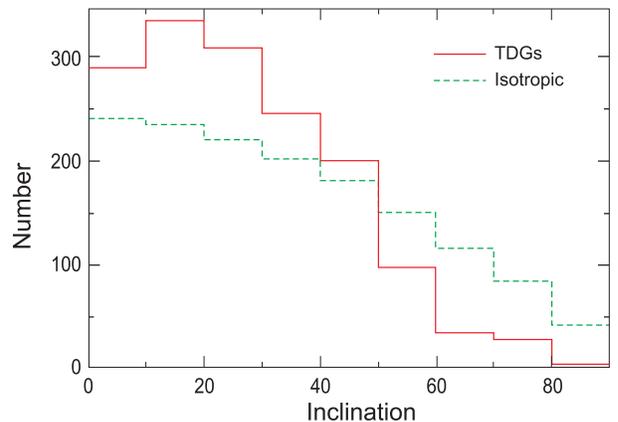}
\caption{Latitudinal distribution of long-lived TDGs around their progenitor galaxies (red), compared to an isotropic distribution (green dashed). TDGs are preferentially distributed towards the equatorial plane of their progenitor disk. This distribution has been computed with data accumulated from $t=500$ to $2000$~Myr, for TDGs that survive at least 1~Gyr. The anisotropy of the distribution increases with radius, as detailed on Fig.\ref{histo_i_4}.}\label{histo_i}
\end{figure}

As shown before, the long-lived tidal objects will behave as satellite galaxies, orbiting around their parent galaxies. We analyze in the following their spatial distribution, as derived from our simulations. To increase the size of the statistical sample, we have cumulated the data on the positions from $t=500$ to $2000$~Myr for objects that survive at least 1~Gyr. The number of ''detections'' is then 1587, corresponding to 143 individual objects tracked at 6 to 16 epochs of their evolution depending of their lifetime. Cumulating data not only enables to increase the statistical size of the sample, but also to make the comparison with samples of real colliding galaxies that are observed at different times of their evolution. Of course, in the case of the simulations, not all the data points are statistically fully independent. 
 
\begin{figure*}[!h]
\centering
\includegraphics[width=16cm]{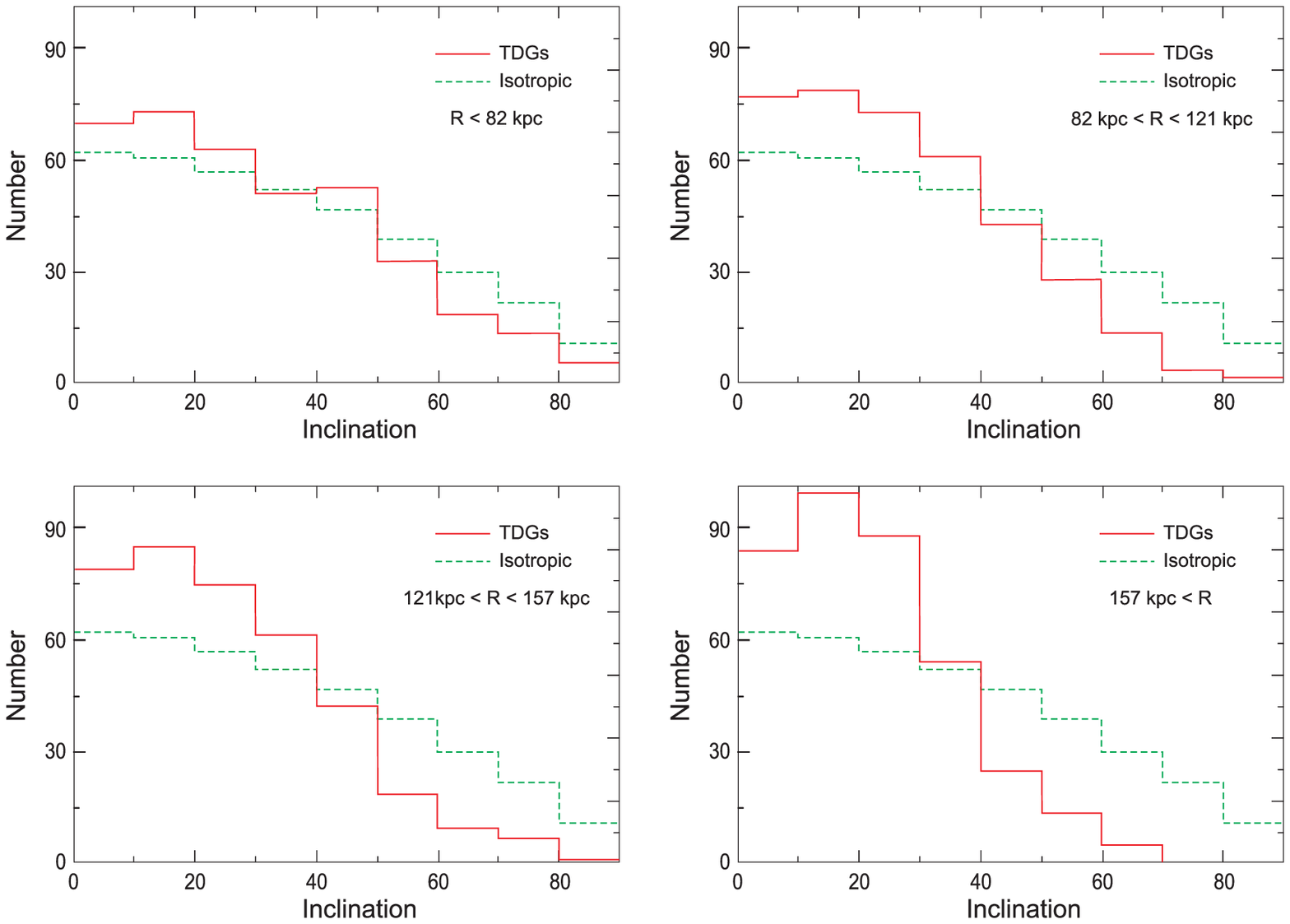}
\caption{Same as Fig.~\ref{histo_i}, but for long-lived TDGs separated into four radial bins of equal population (radial quartiles at 28, 121, and 157~kpc). The anisotropy of the distribution of long-lived TDGs increases with radius.}\label{histo_i_4}
\end{figure*}

\begin{figure*}[!h]
\includegraphics[width=12cm]{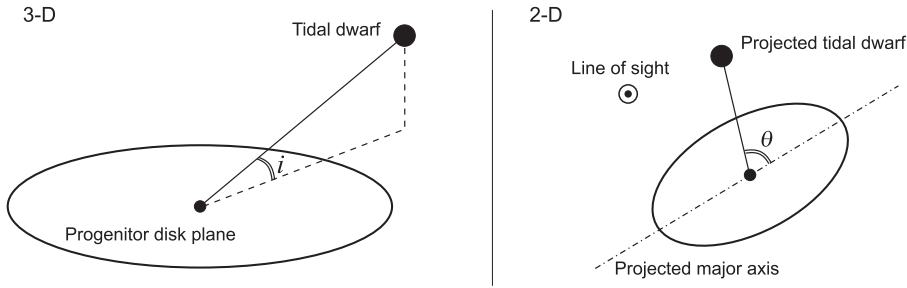}
\caption{Definition of the real inclination of a TDG with respect to the disk plane (angle $i$ in the 3-D space, left), and its apparent position w-r-t the projected major axis of its host (angle $\theta$ in the 2-D projection plane, right). One given object at one given instant has a single latitude $i$ but a probability distribution for $\theta$ varying from 0 to 90 degrees, depending on the line-of-sight. 
}\label{schema}
\end{figure*}

We measure at each time the radius and inclination with respect to the parent galaxy disk. This inclination $i$ is defined as the angle between (i) the parent galaxy disk, or, after the merger, the main flattening plane of the parent galaxy material\footnote{In major mergers forming elliptical galaxies, the main flattening plane of the remnant is often found to correspond to the orbital plane, more than to the parent galaxy plane. However, since TDG-forming cases correspond to low orbital inclinations, the parent galaxy plane and the orbital plane are never very different.}, and (ii) the line linking the parent galaxy mass center to the tidal dwarf mass center (Fig.~\ref{schema}). We show on Fig.~\ref{points_r_i} their distribution in the (radius, inclination) plane. The inclination distribution is shown on Fig.~\ref{histo_i} for all the objects, and on Fig.~\ref{histo_i_4} for the objects separated into four quartiles depending on their initial radius (0--82, 82--121, 121--157 and $>$157~kpc). 

The distribution of the long-lived tidal objects around their parent galaxies (or the associated merger remnants) is strongly anisotropic. The fraction of objects found at galactic latitudes larger than 60 degrees is 4 times smaller than what is expected for an isotropic distribution (Fig.~\ref{histo_i}). Tidal dwarfs are mostly distributed in the equatorial plane of their progenitor disk. As mentioned earlier, they are preferentially formed when the orbital plane and the parent spiral galaxy plane are inclined by less than 40 degrees. In such cases, the tidal tails and the sub-structures formed within them are close to the disk plane. In a few rare cases, interactions with the progenitors and, potentially, with other tidal objects, drive some objects far from the parent disk plane.
 
\begin{figure}[!h]
\centering
\includegraphics[width=8cm]{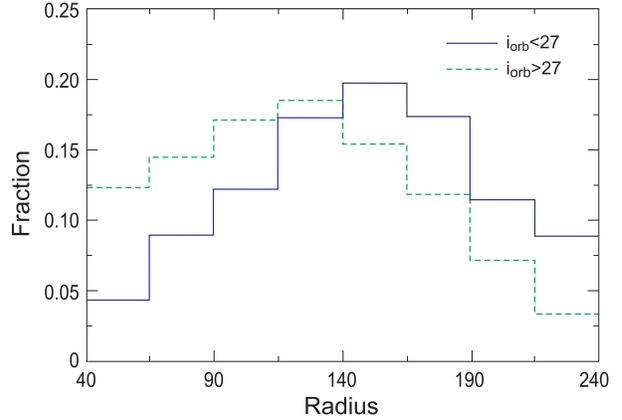}
\caption{Radial distribution of the long-lived objects, depending on the inclination of the companion orbital plane with respect to the progenitor disk plane: orbital inclinations smaller than 27 degrees (blue solid) and larger than 27 degrees (green dashed). Low-inclination orbits form the longest tidal tails and statistically produce TDGs at larger radii than more inclined orbits.}\label{n_r_i}
\end{figure}

The anisotropy of the distribution increases with radius, as seen on Fig.~\ref{histo_i_4}. Indeed, tidal tails are more extended in the close-to-coplanar cases and the objects forming at their tip are located at larger distances (Fig.~\ref{n_r_i}). Finally, the most prominent tidal substructures have a more anisotropic distribution. We measure an average value of the inclination $<i>=31$ degrees for objects less massive than $10^9$~M$_ {\sun}$, and $<i>=22$ 
for the more massive ones. This, again, is explained by the fact that the most massive objects are formed on low orbital inclinations (i.e. nearly-coplanar galactic encounters).
 
\subsubsection{Orbit eccentricities}

\begin{figure}[!h]
\centering
\includegraphics[width=8cm]{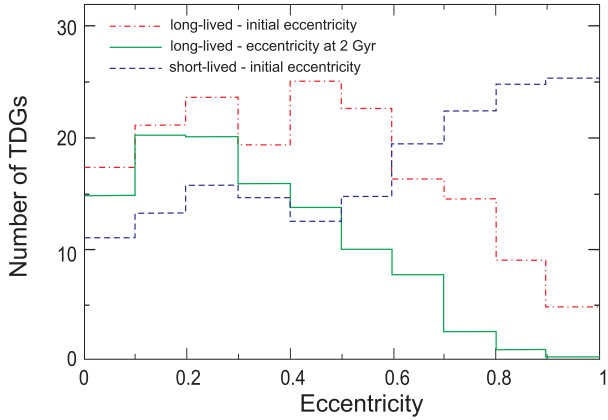}
\caption{Statistical distribution of the initial eccentricity (at $t=200$~Myr) of the objects able to survive at least 1~Gyr which became satellites (red dash-dotted histogram), and final eccentricity for those surviving at least 2~Gyr (green histogram). The dashed blue histogram shows the initial eccentricity of objects surviving less than 1~Gyr, renormalized to the same number of objects than for the red histogram.}\label{eccent}
\end{figure}

We determined the eccentricity of the orbits of the tidal satellites and followed its variation as a function of time. The ''eccentricity'' of the orbit of a given object at a given epoch was defined as follow: the orbit that this object would have if the potential of the merging system did not evolve is computed. A maximal and a minimal radii, respectively $r_+$ and $r_-$ measured from the mass center of the merging/merged spirals, are found. The eccentricity $e$ is then derived assuming that the orbit is in first order elliptical: 
 $e=(r_+ - r_-)/(r_+ + r_-)$. We show in Fig.~\ref{eccent} the distribution of $e$ for the long-lived objects when they are formed, i.e. at their first detection (generally at $t=200$~Myr). Nearly 90\% of them were born on orbits with an axis ratio larger than 0.6; very eccentric orbits are rare. Objects with high eccentricities have a smaller pericenter radius, and are then exposed to a faster and more disruptive tidal field from the central galaxy; they also undergo a higher dynamical friction, causing them to quickly fall back onto their progenitor. At 2 Gyr, they have disappeared. The short-lived tidal structures, also presented in Fig.~\ref{eccent}, had, without surprise, high eccentricities. 
 
We illustrate this point in Fig.~\ref{points_r_i} where we show the orbital track of two objects, initially formed at similar radii, in similar colliding systems (2:1 mergers, run 5 and 6), but with different orbital eccentricities. The most eccentric tidal dwarf is found to undergo a stronger orbital decay, under the effects of dynamical friction. It also loses 18\% of its mass from $t=0.5$ to 2~Gyr, while the less eccentric dwarf loses only 8\% of its mass during the same period. 
 
\subsubsection{Relative velocities}

\begin{figure}[!h]
\centering
\includegraphics[width=8cm]{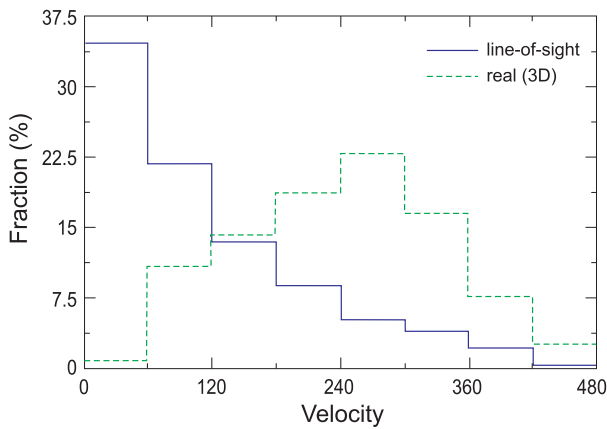}
\caption{Velocity distribution of TDGs with respect to the mass center of the merging pair of massive galaxies: real (3-D) velocity (green dashed) and projected along the line-of-sight, in absolute value (blue solid). Data accumulated from $t=500$ to 2000~Myr, for TDGs surviving at least 1~Gyr.}\label{velo}
\end{figure}

Since they move on low eccentricity orbits, the tidal satellites generally have a velocity close to the circular velocity of the parent gravitational system (dominated by the merged dark halo). The resulting distribution of the velocities at $t=1$~Gyr is shown on Fig.~\ref{velo}. It ranges between 50 and 400~km~s$^{-1}$ (for a massive progenitor with a virial velocity of 220~km~s$^{-1}$). Assuming that systems are observed under random orientation, we computed the distribution of the projected line-of-sight velocities (see Fig.~\ref{velo}). The average absolute value is $< |V_{\mathrm{LOS}}| > =115$~km~s$^{-1}$ with respect to the mass center of the central merger remnant. In the case of young systems, a few $10^8$~yr after the merger, tidal tails are more easily seen edge-on \citep[e.g.][]{HB04} so that the real line-of-sight distribution is not isotropic. This can statistically decrease the projection angle, which in turns increases the observed line-of-sight velocities. The observed distribution around young systems should then fall somewhere in between the two histograms shown on Fig.~\ref{velo}.
 Still, typical observed velocities will rarely exceed the order of 200-250~km~s$^{-1}$.

\subsubsection{Dark matter content}\label{DMfrac}

We have measured the amount of dark matter inside the visible radius\footnote{defined here as the radius containing 90\% of the baryonic mass (stars+gas) of the object} for tidal objects that survive more than 1~Gyr. We find that inside this radius, the dark-to-visible mass ratio is on the average 0.09. It is never more than 0.15, with the highest dark matter fractions generally found in the most massive TDGs. For comparison, massive spiral progenitors in our initial conditions have a dark-to-visible mass ratio of 0.65 inside their visible radius.

Hence, tidal dwarf galaxies are strongly deficient in dark matter, according to our simulations. \citet{BH92} found a similar result; our tidal dwarfs contain slightly more dark matter because we use more extended dark halos around the massive progenitors, so that more dark matter is found in the vicinity of forming TDGs.

\section{Discussion: comparison to observed TDGs and satellite galaxies}
\label{obs}

\subsection{Identifying young forming Tidal Dwarf Galaxies in colliding galaxies}\label{obs1}

A large variety of star-forming entities have been reported in the tidal tails of real interacting systems: 
Super Star Cluster (SSCs), Giant HII Complexes (GHCs) \citep{gallagher,knierman,weilbacher1,weilbacher2,saviane,LS04,IP01,sakai} and the more massive so-called "Tidal Dwarf Galaxies candidates" which, in fact, also correspond to a diversity of objects. 
 However, several observed interacting systems are distinguished by the massive accumulations of matter, up to a few $10^9$~M$_{\sun}$, found at the tip of their tidal tails \citep[e.g.][]{DM94,DM98,HB04}. In principle, they could be fake condensations caused by the projection of material along the line-of-sight in tidal tails seen edge-on. However their molecular gas content \citep{braine} and their large-scale kinematics provide evidences that they are real massive objects, while their internal kinematics, when resolved, indicates that these structures are gravitationally bound \citep{bournaud04}. Most of these tidal objects are observed at a time when they are probably still forming, a few hundred Myr after the beginning of the collision. Without any detailed observations and the help of evolutionary models, it is difficult to predict how they will evolve, and in particular whether they will become independent, stable, objects that would deserve to be called Tidal Dwarf Galaxies. Our large set of simulations show that among the objects formed at the base and the middle of the tidal tails, many are on rather high eccentric orbits which hampers considerably their ability to survive. Actually only a small fraction of them are still present after 500~Myr. About 75 percent of the objects more massive than $10^8$~M$_{\sun}$ -- our mass threshold -- survive less than 800~Myr. Such objects may be considered as TDG-candidates in on-going mergers, but will most likely be short-lived and not become dwarf galaxy satellites orbiting around the merger remnant for at least a few dynamical times. \citet{kroupa97} studied the evolution of purely stellar low-mass satellite galaxies without dark matter (as expected for TDGs) in the potential well of the Galactic dark halo. He found that satellites on high eccentric orbits lose 10--20 \% of their mass during each perigalactic passage. They nevertheless manage to survive reaching a stationary regime but their remnants contain then only 1 \% of their initial mass, about $10^5$~M$_{\sun}$ in the simulations of \citet{kroupa97}. To have final masses typical of dwarf spheroidals -- the less massive known galaxies --, they then should initially have had stellar masses as high as $10^9 - 10^{10} $~M$_{\sun}$, which can correspond only to the most massive objects, formed at the tip of tidal tails in our simulations. 
 
According to our simulations, the only objects that actually become genuine dwarf galaxies are those formed at large radii, and preferentially near the extremity of the tails (Fig.~\ref{dist_rad}) via the mechanism studied by D04 which works for extended dark matter haloes. We then conclude that among the many structures observed in the tidal tails of young interacting systems, only the most massive ones that are formed near the tip of tails are likely to be the progenitors of Tidal Dwarf Galaxies, following the definition given e.g. by D04 or \citet{HB04} that requires them to be long-lived. 

One should however keep in mind the restrictions of these simulations. The produced TDGs were found to be long-lived essentially because they were orbiting on safe orbits (at large radii and low-eccentricities) and were not affected by a strong tidal field that could otherwise disrupt them. However we have not considered here the internal processes that could cause severe disturbances. Intensive feedback following an internal starburst may remove much gas from the TDGs \citep{dekel03}. Massive objects may also fragment into smaller ones in higher resolution models. We will present in a separate paper high resolution simulations zooming in onto TDGs that allowed us to conclude that, even taking into account their internal evolution, they are still long-lived objects that conserve most of their initial mass.

\subsection{Identifying Tidal Dwarf Galaxies around merger remnants}\label{obs1b}

Young forming Tidal Dwarf Galaxies are rather easily pinpointed as long as they are still linked to their parent galaxies by a tidal tail. As explained earlier, an HI accumulation or CO detection at its tip make a it a promising TDG candidate. However, once they become "isolated" and that the umbilical cord is broken -- typically tidal tails dissolve in 300--500 Myr --, they become much more difficult to identify. Indeed, they resemble satellite galaxies orbiting the merger remnant.

Several criteria to establish a tidal origin were discussed in the literature:
an unusually high metallicity received in in-heritage from their parent galaxies \citep{DM98}, a deficiency of non--baryonic dark matter \citep{BH92} and thus a particular location in the Tully-Fischer diagram \citep{hunter}. Several quests for identifying old TDGs among the dwarf galaxy population were carried out (see \citet{duc04iau} and references therein), but sofar no dwarf satellite formed more than 1~Gyr ago during a collision has ever been unambiguously recognized, partly because the above mentioned criteria are often difficult to check.
 \smallskip

The simulations presented here provide further and somehow simpler constraints to probe a tidal origin:

\paragraph{Mass criterion}
The mass of TDGs is typically 0.1 to 1\% the baryonic (gas+star) mass of their progenitor when the latter had a gas content typical of bright spiral galaxies before the merger. It can be at the very most a few percent in the case of gas-rich progenitors. Even more massive tidal galaxies were only able to form at high redshift when the parent spiral galaxies contained more gas.
It is likely that such objects have already disappeared at redshift 0. They would be anyway difficult to identify from their metallicity which is not expected to be very high. We can apply the mass criterion to some nearby systems (though still young) where the presence of massive TDGs had been suspected but later debated. For instance, the northern TDG-candidate of Arp~105 \citep{duc97} has a mass of nearly 10 percent of the parent galaxy mass ($7\times10^9$ solar masses in HI, the dominant component). An object with such a mass is then unlikely to be a real tidal dwarf, unless its mass had been over-estimated due, for instance, to the contribution of projected foreground or background material along the line-of-sight. The presence of such a projection effect in the northern tail of Arp~105 is indeed suggested by the large-sale kinematics of the tail \citep{bournaud04}. The Southern TDG-candidate identified by \citet{duc97}, with a mass of 0.5\% of its parent spiral, is much more likely to be a genuine tidal dwarf.

This mass criterion also suggests that the potential TDG-candidate NGC~3077 in the group of M~81, linked to M~81 by a gaseous bridge \citep{yun94} is unlikely to be of tidal origin: according to its infrared luminosity and estimated HI mass, the mass of this galaxy is indeed as high as 15--20\% of the mass of M~81.

\paragraph{Spatial distribution}
Statistically, the simulated TDGs tend to be concentrated towards the equatorial plane of their massive host galaxy. However, this property is not a strong one for individual cases, since one given TDG can be
 found in a polar plane (see Fig.~\ref{histo_i}). Moreover, in the case of equal-mass mergers, the equatorial plane of the spiral progenitor may hardly be observationally determined. Additionally, TDGs are rarely found at more than ~15 times the optical radius of their progenitor, i.e. typically 200-250 kpc around a bright early-type host galaxy. In compact galaxy groups, the interaction with other galaxies may kick them from the host potential, and TDGs may be found further out from their progenitor.

\paragraph{Velocity}
The relative velocities of TDGs, with respect to their hosts, are of the order of the circular velocities in dark haloes. Projected along the line-of-sight, this rarely results in velocities larger than 200--250~km~s$^{-1}$. Higher-velocity dwarf satellites are unlikely to be of tidal origin.

\paragraph{Dark matter content}
Dwarf galaxies of tidal origin are expected to be deficient in dark matter (see Sect.~\ref{DMfrac} and \citet{BH92}). Their dark matter content is negligible compared to their progenitor and to the many
 dwarfs with measured high M/L ratios.
 Note however that this prediction is made within a pure CDM frame. The presence of a baryonic and dissipative dark matter component \citep[e.g.][]{pfenniger} could radically change the result, for this component could participate to the formation of TDGs. 

\paragraph{Properties of the host galaxy}
Long-lived TDGs are formed during galaxy collisions in the range of mass ratio 1:1 to 8:1, and more efficiently between 1:1 and 4:1. According to recent simulations \citep{bournaud05,naab06}, 
the remnant of such barely-equal mass merger should have the morphology of an Sa or more likely an S0 or even that of an elliptical, unless the colliding galaxies were particularly gas rich, as in the Early Universe. Tidal satellites should then be found more frequently around early-type galaxies.

We did not explore parameters for which the parent galaxies do not merge and hence may remain spirals. In such fly-bys, which generally correspond to large impact parameters, the corotation radius is large; less material gets into tidal tails and the latter are shorter. If TDGs are produced at all, they will have lower masses. Indeed, the high-velocity or large-distance cases that we have simulated (that are the closest to fly-bys without mergers), correspond to less numerous and less massive tidal dwarfs than the average (see Appendix~A).

\smallskip

\subsection{Satellite galaxies around their hosts}\label{anis2D}\label {obs2}

We have shown in the Section~3 that the distribution of TDGs around their progenitor galaxies is anisotropic. However, the projection on the sky may hide this property. Indeed, a TDG that in the 3D space lies in the equatorial plane of its massive progenitor, can be observed in the 2D projection aligned with the minor axis of its massive host, causing possible confusion with a dwarf satellite on a polar orbit.

\begin{figure}[!h]
\centering
\includegraphics[width=8cm]{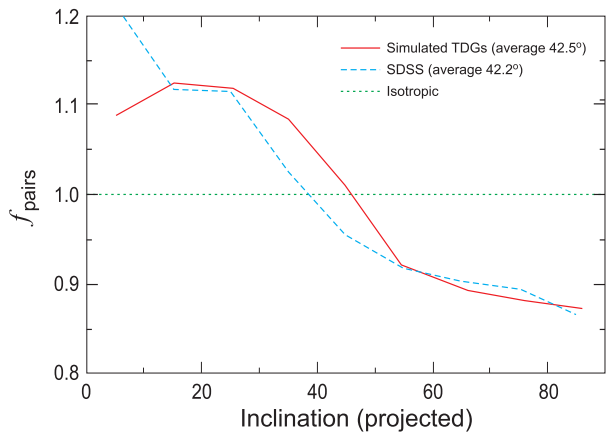}
\caption{Distribution of the projected inclination of dwarf satellites of tidal origin w.r.t. their host (progenitors) galaxy plane: the angle $\theta$ between the dwarf galaxy and the host galaxy apparent major axis is measured (see Fig.~\ref{schema}). The distribution has been derived assuming that the systems are randomly projected on the sky. The average value of $\theta$ is 42.5 degrees, while an isotropic distribution corresponds to $\theta=45$ degrees. The slightly more anisotropic distribution observed in the Sloan Digitized Sky Survey by Yang et al. (2006) is also shown.}\label{aniso2D}
\end{figure}

We have statistically estimated the apparent angle $\theta$ between the host galaxy major axis and the apparent position of the TDG, defined on Fig.~\ref{schema}. This is the usual observational method used to study the distribution of dwarf satellites around bright galaxies \citep[e.g.][and references therein]{yang06}. To infer the distribution of $\theta$ from that of the latitude $i$ in three dimensions, we assumed that the host+TDG systems are observed under isotropically distributed line-of-sights. The calculation was made analytically rather than assuming a limited number of possible line-of-sights in simulations. The result is shown of Fig.~\ref{aniso2D}~: after projection on the sky plane, the anisotropy in the distribution of TDGs is still visible. 

Most studies of the distribution of dwarf satellites around massive host galaxies focus on the most massive dwarfs. For instance, the work of \citet{brainerd05} based on SDSS data includes only satellites more massive than 1/8 of their host mass. Such objects are unlikely to be of tidal origin, since they are 1--2 orders of magnitude above the typical masses of long-lived TDGs. A study including lower mass objects has been performed by \citet{yang06}. These authors found an anisotropic distribution of dwarf satellites, similar to what is found the simulations tidal dwarfs. These results cannot be directly compared, since certainly not all the dwarfs are of tidal origin (see next section). However, TDGs may significantly contribute to forster the observed anisotropy in the distribution of dwarf satellites. Hence, the anisotropy of dwarf satellites may not only relate to the shape of dark haloes, but also to the tidal origin of some satellites. Therefore an anisotropic distribution could in principle be present even within spherical dark halos. Also note that \citet{yang06} find that the distribution of satellites is more anisotropic around red, early-type host galaxies. This observational fact could be accounted for by the anisotropy of TDGs since tidal dwarfs are more likely to be found around early-type hosts.

\subsection{Contributions of TDGs to the dwarf population}\label{fractdg}

The rate of TDG formation in galaxy mergers cannot be directly deduced from our set of simulations, for we deliberately privileged favorable cases: the distribution of mass ratios and orbital parameters in our simulations is not aimed at being cosmologically realistic. However, we have derived the constraints on parameters for the formation of long-lived tidal dwarfs (Section~3.2) and this can be used to estimate the actual frequency at which TDGs are formed.

Our simulations indicate that on average a bright spiral galaxy produces 1.3 tidal dwarf (considering only those that survive at least 1 Gyr), whenever it undergoes a merger with a galaxy between 1/4 and 8 times its mass, on a prograde orbit, with an orbital plane inclined up to 40 degrees with respect to its disk plane. Assuming that real orbits for galactic encounters have an isotropic distribution, this corresponds to a production rate of 0.4 TDG per merging spiral. For mergers in the 1:1--4:1 range of mass ratios, both spirals can form long-lived TDGs, and the production rate increases to 0.8 TDG per merging pair. 

This estimate {\it a priori} assumes that the distribution of impact parameters and initial velocities in our sample of simulation is statistically representative of real colliding systems, which is probably not the case. However, our results summarized in Table~1 show these two parameters have a minor influence on the formation of TDGs. Then, even if their distribution in our sample is not realistic, this does not result in an important bias on the estimated rate of TDG production.

\citet{okazaki} claimed that all dwarfs in the Universe could have a tidal origin. Their estimate relied however on a production rate of 1 to 2 tidal dwarfs having a live time of at least 10 Gyr, which is actually much higher than that predicted by our simulations.

The number of TDGs versus time in our simulations (207 at $t=500$~Myr, 143 at $t=1$~Gyr, and 119 at $t=2$~Gyr) fits well with an exponential decay and a lifetime of 2.5~Gyr, so that only 20 percent of the TDGs actually survive up to 10~Gyr. Therefore, our best production rate -- 0.8 TDG with life of 1 Gyr per merging pair -- corresponds to just 0.1--0.2 TDG with a life of 10~Gyr, ten times lower than the value assumed by \citet{okazaki}. Using their other hypotheses, in particular the evolution of the number of mergers with redshift, we conclude that at most 10\% of the dwarf galaxy population could be of tidal origin. 
 
In fact, the study by \citet{okazaki} implicitly covers a larger range of mass ratios. Only about one third of the mergers are likely to fall within the 1:1--8:1 range favorable to long-lived TDGs. Since the latter estimate is very tentative, we will simply conclude that a few percent of the dwarf galaxy population can be of tidal origin. On the other hand, we did not take into account the fact that more distant galaxies were more gas rich and had probably an HI disk much more extended than in local spirals, two properties which may enhance the TDG production rate.
 
Only few observational studies have sofar tried to quantify the contribution of old, "detached" (thus long--lived) TDGs to the dwarf population. \citet{DD03} did not found any excess of dwarfs around a sample of strongly interacting systems and concluded that the contribution of long-lived TDGs should be rather minor. On the other hand, in denser environments, multiple collisions could kick tidal dwarfs from the progenitor potential and thus contribute to increase their life expectancy. \citet{Hunsberger96} probed the dwarf population of a sample of compact groups of galaxies and concluded that 50\% of them were probably tidal objects. The production of tidal dwarfs in the cluster environment could be hampered by the fact that tidal tails develop in a less efficient way and tend to be more diffuse in massive potential wells \citep{Mihos-IAUS217}. However such objects may have formed in sub-groups falling in the cluster. Arp~105 \citep{DM94}, NGC~5291 \citep{DM98}, IC~1182 \citep{Iglesias-Paramo03} are instances of systems belonging to clusters or sub-structures within clusters where young TDG candidates have actually been identified. The presence of a large population of cluster dwarf ellipticals in Coma and Fornax \citep{Poggianti01,Rakos00}, of dwarf irregulars in Hydra-I and Hercules \citep{Duc01,Iglesias-Paramo03} having deviant -- high -- metallicities for their luminosity could be interpreted  as hints for a tidal origin for some of them although other mechanisms could explain these anomalies \citep{Conselice03}.

\subsection{Looking for TDGs in the Local Group}\label{obs3}

The distribution of the dwarf satellites of the Milky-Way is observed to be anisotropic within so-called Great Plane(s) \citep{LB76,K79,Maj94}. This fact was exploited by \citet{kroupa05} and \citet{sawa} to claim that the dSphs around the MW could be of tidal origin. However, the satellites are actually not concentrated towards the disk plane, but along one or two nearly perpendicular directions. Our simulations indicate that such a distribution is a priori in conflict with the hypothesis that the M-W dSphs are old TDGs since the latter are statistically concentrated towards the equatorial plane of their progenitor\footnote{though high latitudes TDGs can also be present; see Fig.\ref{histo_i}} ... unless the parent galaxy is not the M-W. Let us imagine that long ago a spiral with a disk perpendicular to the M-W disk, orbited in a polar plane w.r.t the M-W and produced in that plane long-lived TDGs. Nowadays, we do not see the disk of this progenitor, either because it has flown far away or it has merged with the M-W disk. According to our simulations, the large number of putative TDGs -- more than 10 dSphs are orbiting around the M-W -- indicate that either the collision at their origin occurred less than 1~Gyr ago, before most of them are destroyed -- however, no sign of such recent encounter is clearly visible, except the LMC/SMC/interaction --, or that several mergers, all with specific orbits included in the Great Plane(s), occurred long ago and produced each 1--2 long--lived TDGs. However, the present morphology of the Galaxy is inconsistent with the idea that it has merged in the past with several companions with mass ratios of at least 8:1, the limits we found to produce a TDG: if this had occurred, the M-W would have become an S0 or E galaxy \citep[e.g.][]{bournaud05}.

In that respect, M~31 is more likely to host tidal satellites; its morphology has an earlier type and numerous signs of possible past interactions, such as long stellar streams, were found in its outer most regions \citep{ibata01}. \citet{koch06} discuss the three-dimensional location of the satellite galaxies around M~31. The global distribution of satellites in this system do not show any clear concentration towards a particular plane, but the early-type dwarfs and the dSphs in particular seem to be concentrated towards a polar plane. As for the case of the Milky-Way, this distribution would require that several galaxy mergers have occurred on this configuration, which makes the tidal scenario for the origin of M~31's dwarfs unlikely. 

Therefore, just from the distribution and number of the dSphs located in the vicinity of the M-W, and the statistical results obtained
with our simulations, one should conclude that it is unlikely that the majority of them are of tidal origin. This is also the case for M~31. Other scenarios to explain their distribution in great planes \citep[e.g.][]{koch06,LK06} are then more realistic.

However, other, somehow more direct, criteria to probe a tidal origin should be investigated. For instance, a deviation from the luminosity--metallicity relation \citep{duc2000} due to pre-enrichment processes in the parent's gaseous disk or the absence of a non baryonic dark matter component (see Sect.~\ref{DMfrac}) are clear characteristics of TDGs.

In that respect, the Local Group dwarfs were subject of active debates. They are usually considered as being the objects showing the highest M/L ratios (up to 100) and thus containing the highest quantities of dark matter, which would argue against a tidal origin, at least in a CDM context. Nevertheless the accurateness of the estimate of their total mass is controversial, and claims were made that it had been largely over-estimated due to the lack of statistics -- the dynamical mass is derived from the velocity dispersion of a limited number of individual stars -- or due to the tidal shaking of the stars \citep{kroupa05}. Beside, the Local Group dwarfs are remarkable for their variety of M/L ratios, chemical enrichment and more globally star formation history despite being all in the same environment \citep{grebel01}. This leaves room for a tidal origin in some of them. 

\subsection{A note on merger induced globular clusters}
In our analysis, we only considered sub-structures with masses exceeding $10^8$~M$_{\sun}$ to make sure they contained enough particles and had a fair resolution. Our numerical study did not aim at making predictions on lower mass objects, in particular the potentially forming globular clusters\footnote{Note though on Fig.~1 the presence of low mass objects in the vicinity of the merger remnant.}. Objects of masses typically $10^5-10^7$~M$_{\sun}$ could be numerous in tidal tails, and their survival less problematic than the objects with typical masses of $10^8$~M$_{\sun}$. Indeed, if concentrated enough, they will be less affected by tidal disruption, undergo less dynamical friction and face a slower orbital decay \citep{miocchi06}. \citet{kroupa97} studied the evolution of $10^5$~M$_{\sun}$ dark matter free satellites and concluded that they could be long-lived, as well as globular clusters possibly formed in mergers. Higher resolution simulations including a description of the formation process are needed to study this scenario of globular cluster formation in mergers.


\section{Conclusion}

In this paper, we have studied the formation and the evolution of Tidal Dwarf Galaxies and related candidates, in about 100 numerical simulations of collisions of spiral galaxies, with gas content typical of gas-rich low to moderate redshift spirals. The identification of massive ($M > 10^8$~M$_{\sun}$) substructures formed in tidal tails, and their follow-up during 2~Gyr, have shown that:

\begin{itemize}

\item Among the numerous tidal objects, those susceptible to survive more than 1~Gyr were formed in the outer parts of tidal tails; they also turn out to be the most massive ones. Gravitational clumps which grow all along the tails rarely survive more than 500~Myr, mainly because they are formed on more eccentric orbits, and generally fall back onto their progenitors or are disrupted by the tidal field losing much of their initial mass. These objects should not be considered as genuine Tidal Dwarf Galaxies, according to the definition of \citet{duc04iau} or \citet{HB04}, even if they have, at the moment they are observed, the apparent characteristics of young forming dwarf galaxies.

\item Long-lived tidal dwarf galaxies are preferentially formed on prograde, close-to-coplanar orbits (inclination smaller than $\sim$ 40 degrees). In such cases, more than 2 long-lived TDGs are rarely formed from one progenitor galaxy. These results have led us to the conclusion that a small but significant fraction of a few percent of dwarf satellites could be of tidal origin. The fraction of tidal dwarfs can be larger around early-type hosts that have undergone numerous mergers or in rich environments where tidal dwarfs can be longer-lived.

\item The statistical properties of the long-lived TDGs, those that become dwarf satellites, have been studied. They are typically found within 10--15 optical radii from their progenitor, with moderate relative velocities, and masses at the very most a few percent of their progenitor mass (or about 10\% of their progenitor initial gaseous mass). Their spatial distribution is anisotropic, flattened towards the equatorial plane of the parent spiral galaxy. Would the contribution of TDGs be important in some environments, they would then contribute to foster the anisotropy already observed for the distribution of the general dwarf satellite population. 

\item Given their number and spatial distribution in polar Great Planes, the majority of the satellites of the Milky-Way and M~31 are unlikely to all be of tidal origin. 

\end{itemize}

These results can help identifying, in real young interacting systems, the tidal objects that are the most likely to be long-lived and provide hints on the location of the long searched ''old'' TDGs which, once the tidal tails linking them to their parents have disappeared, are difficult to identify. If searched randomly without any selection criteria, tens of dwarf satellites would have to be analyzed for a few TDGs to be found. 

Galaxy mergers can also produce a significant number of smaller mass objects like globular clusters, but the resolution of the present simulations did not enable us to study them in detail. Also, the internal properties of TDGs have not been studied in this paper. They will be the subject of a forthcoming study using higher resolution simulations, dedicated to resolve the internal dynamics of tidal substructures, in particular those that are now known to be long-lived.


\begin{acknowledgements}
We are grateful to an anonymous referee for a careful reading of the manuscript, and to Fran\c{c}oise Combes for her comments on a previous version. The numerical simulations were carried out on the NEC-SX6 vectorial computer of the CEA/CCRT, and we are most grateful to Frederic Masset for helping with access to this computing center. This research was led within the Horizon project (\texttt{http://www.projet-horizon.fr}). 
\end{acknowledgements}



\appendix

\section{Parameters and results of numerical runs}\label{appen}

We give in Table~\ref{table_run} the parameters used for each simulation. The mass ratio is $M$. The orientation of the orbit, prograde (P) or retrograde (R) is given for the most massive galaxy and the less massive one respectively. The impact parameter $R$ and relative velocity at an infinite distance $V$ are computed assuming that dynamical friction is negligible before the beginning of the simulation. The initial inclination of the orbital plane to the disk plane is given by $i_1$ and $i_2$ for each colliding galaxy. $N_1$ and $N_2$ are the number of long-lived tidal dwarf galaxies (surviving at least 1~Gyr) more massive than $10^8$~M$_{\sun}$ formed in the material of each parent galaxy. Galaxy number 1 is the most massive one.

\begin{table*}
\centering
\begin{tabular}{lccccccccc}
\hline
\hline
Run & M & Orient. & $R$ & $V$ & $i_1$ & $i_2$ & & $N_1$ & $N_2$ \\ 
\hline
0 & 1 & P/P & 4 & 150 & 0 & 0 & &2& 2 \\ %
1 & 1 & R/R & 4 & 150 & 0 & 0 & &-& - \\ %
2 & 1 & P/R & 4 & 150 & 0 & 0 & &2& - \\ %
3 & 2 & P/P & 4 & 150 & 0 & 0 & &1& 1 \\ %
4 & 2 & R/R & 4 & 150 & 0 & 0 & &-& - \\ %
5 & 2 & P/R & 4 & 150 & 0 & 0 & &1& - \\ %
6 & 5 & P/P & 4 & 150 & 0 & 0 & &1& 1 \\ %
7 & 5 & R/P & 4 & 150 & 0 & 0 & &-& 2 \\ %
8 & 2 & P/P & 4 & 150 & 35 & 35 & &1& 1 \\ %
9 & 2 & P/R & 4 & 150 & 35 & 35 & &2& - \\ %
10 & 2 & P/P & 4 & 150 & 15 & 15 & &1& 2 \\ %
11 & 2 & P/P & 4 & 150 & 45 & 45 & &1& - \\ %
12 & 2 & P/P & 4 & 150 & 60 & 60 & &-& - \\ %
13 & 5 & P/P & 4 & 150 & 35 & 35 & &-& 2 \\ %
14 & 5 & P/R & 4 & 150 & 35 & 35 & &-& - \\ %
15 & 5 & P/P & 4 & 150 & 15 & 15 & &1& 2 \\ %
16 & 5 & P/P & 4 & 150 & 45 & 45 & &-& - \\ %
17 & 5 & P/P & 4 & 150 & 60 & 60 & &-& - \\ %
18 & 2 & P/P & 1 & 150 & 30 & 30 & &-& 1 \\ %
19 & 2 & P/P & 2 & 150 & 30 & 30 & &-& 3 \\ %
20 & 2 & P/P & 6 & 150 & 30 & 30 & &1& 1 \\ %
21 & 2 & P/P & 9 & 150 & 30 & 30 & &1& 2 \\ %
22 & 2 & P/P & 1 & 80 & 30 & 30 & &-& 1 \\ %
23 & 2 & P/P & 2 & 80 & 30 & 30 & &1& 2 \\ %
24 & 2 & P/P & 6 & 80 & 30 & 30 & &1& 2 \\ %
25 & 2 & P/P & 9 & 80 & 30 & 30 & &1& 2 \\ %
26 & 2 & P/P & 12 & 80 & 30 & 30 & &2& 1 \\ %
27 & 2 & P/P & 7 & 30 & 30 & 30 & &-& - \\ %
28 & 2 & P/P & 7 & 50 & 30 & 30 & &-& 2 \\ %
29 & 2 & P/P & 7 & 80 & 30 & 30 & &1& 1 \\ %
30 & 2 & P/P & 7 & 110 & 30 & 30 & &1& 2 \\ %
31 & 2 & P/P & 7 & 150 & 30 & 30 & &1& 1 \\ %
32 & 2 & P/P & 7 & 190 & 30 & 30 & &1& 2 \\ %
33 & 2 & P/P & 7 & 250 & 30 & 30 & &1& 1 \\ %
34 & 2 & P/P & 7 & 320 & 30 & 30 & &-& - \\ %
35 & 5 & P/P & 7 & 30 & 30 & 30 & &-& 1 \\ %
36 & 5 & P/P & 7 & 50 & 30 & 30 & &-& 2 \\ %
37 & 5 & P/P & 7 & 80 & 30 & 30 & &-& 1 \\ %
38 & 5 & P/P & 7 & 110 & 30 & 30 & &-& 1 \\ %
39 & 5 & P/P & 7 & 150 & 30 & 30 & &-& 2 \\ %
40 & 5 & P/P & 7 & 190 & 30 & 30 & &-& 1 \\ %
41 & 5 & P/P & 7 & 250 & 30 & 30 & &-& 1 \\ %
42 & 5 & P/P & 7 & 320 & 30 & 30 & &-& 1 \\ %
43 & 5 & P/P & 1 & 150 & 30 & 30 & &-& 1 \\ %
44 & 5 & P/P & 2 & 150 & 30 & 30 & &-& 1 \\ %
45 & 5 & P/P & 6 & 150 & 30 & 30 & &-& 1 \\ %
46 & 5 & P/P & 9 & 150 & 30 & 30 & &1& 2 \\ %
47 & 5 & P/P & 1 & 80 & 30 & 30 & &-& - \\ %
48 & 5 & P/P & 2 & 80 & 30 & 30 & &-& 1 \\ %
49 & 5 & P/P & 6 & 80 & 30 & 30 & &1& 1 \\ %
50 & 5 & P/P & 9 & 80 & 30 & 30 & &-& 2 \\ %
51 & 5 & P/P & 12 & 80 & 30 & 30& &-& 1 \\ %
52 & 8 & R/R & 5 & 70 & 0 & 0 & &-& - \\ %
53 & 8 & R/R & 8 & 70 & 0 & 0 & &-& 1 \\ %
53 & 8 & R/R & 12 & 70 & 0 & 0 & &-& 1 \\ %
54 & 8 & R/R & 5 & 220 & 0 & 0 & &-& - \\ %
55 & 8 & R/R & 8 & 220 & 0 & 0 & &-& - \\ %
\hline
\end{tabular}
\caption{Run parameters and results: number of long-lived TDGs (surviving at least up to $t=1$~Gyr) formed in the material of each parent galaxy. Parameters are defined in text.}
\label{table_run}
\end{table*}

\begin{table*}
\centering
\begin{tabular}{lccccccccc}
\hline
\hline
Run & M & Orient. & $R$ & $V$ & $i_1$ & $i_2$ & & $N_1$ & $N_2$ \\ 
\hline
56 & 2 & P/P & 10 & 60 & 0 & 0 & &1& 2 \\ %
57 & 2 & P/P & 10 & 220 & 0 & 0 & &1& 1 \\ %
58 & 2 & P/P & 10 & 60 & 45 & 45 & &-& - \\ %
59 & 2 & P/P & 10 & 220 & 45 & 45 & &-& - \\ %
60 & 4 & P/P & 10 & 60 & 0 & 0 & &1& 1 \\ %
61 & 4 & P/P & 10 & 220 & 0 & 0 & &1& 2 \\ %
62 & 4 & P/P & 10 & 60 & 45 & 45 & &-& 1 \\ %
63 & 4 & P/P & 10 & 220 & 45 & 45 & &-& - \\ %
64 & 2 & P/P & 2 & 60 & 0 & 0 & &1& 1 \\ %
65 & 2 & P/P & 2 & 250 & 0 & 0 & &-& - \\ %
66 & 2 & P/P & 2 & 60 & 45 & 45 & & -& 1 \\ %
67 & 2 & P/P & 2 & 250 & 45 & 45 & &-& - \\ %
68 & 1 & P/P & 5 & 70 & 0 & 0 & &1& 2 \\ %
69 & 1 & P/P & 5 & 250 & 0 & 0 & &1& 1 \\ %
70 & 1 & P/P & 5 & 70 & 45 & 45 & &-& 1 \\ %
71 & 1 & P/P & 5 & 250 & 45 & 45 & &-& - \\ %
72 & 1 & P/P & 5 & 250 & 40 & 40 & &-& - \\ %
73 & 2 & P/P & 5 & 250 & 40 & 40 & &-& - \\ %
74 & 3 & P/P & 5 & 250 & 40 & 40 & &-& 1 \\ %
75 & 3 & P/P & 1 & 170 & 30 & 30 & &-& - \\ %
76 & 3 & P/P & 2 & 170 & 30 & 30 & &-& 1 \\ %
77 & 3 & P/P & 6 & 170 & 30 & 30 & &2& 1 \\ %
78 & 3 & P/P & 9 & 170 & 30 & 30 & &1& 1 \\ %
79 & 3 & P/P & 1 & 90 & 30 & 30 & &-& 1 \\ %
80 & 3 & P/P & 2 & 90 & 30 & 30 & &-& 1 \\ %
81 & 3 & P/P & 6 & 90 & 30 & 30 & &1& 2 \\ %
82 & 3 & P/P & 9 & 90 & 30 & 30 & &1& 1 \\ %
83 & 3 & P/P & 12 & 90 & 30 & 30 & &1& 1 \\ %
84 & 1 & P/P & 1 & 170 & 10 & 20 & &-& - \\ %
85 & 1 & P/P & 2 & 170 & 10 & 20 & &1& 1 \\ %
86 & 1 & P/P & 6 & 170 & 10 & 20 & &1& 1 \\ %
87 & 1 & P/P & 9 & 170 & 10 & 20 & &1& 2 \\ %
88 & 1 & P/P & 1 & 90 & 10 & 20 & &2& 1 \\ %
89 & 1 & P/P & 2 & 90 & 10 & 20 & &1& 1 \\ %
90 & 1 & P/P & 6 & 90 & 10 & 20 & &2& 2 \\ %
91 & 1 & P/P & 9 & 90 & 10 & 20 & &1& 2 \\ %
92 & 1 & P/P & 12 & 90 & 10 & 20 & & 1& 1 \\ %
93 & 10 & P/P & 4 & 120 & 10 & 10 & &-& - \\ %
94 & 10 & P/P & 8 & 120 & 10 & 10 & &-& - \\ %
95 & 10 & P/P & 6 & 180 & 10 & 10 & &-& - \\ %

\hline
\end{tabular}
\caption{Run parameters (continued).}
\label{table_run2}
\end{table*}


\end{document}